\documentclass{jaa}
\usepackage{natbib}
\usepackage{booktabs}
\usepackage{siunitx}
\usepackage{graphicx}

\newcommand{\Hei}{\ensuremath{\textrm{He}\,\textsc{i}}}

\begin{document}\sloppy

%%paper title
%%For line breaks \\ can be used within title
%\title{Shock-Excited Molecular Hydrogen in the Orion Protostar HOPS\,370}
\title{Unveiling the Shock-Excited Molecular Environment in the Orion Protostar HOPS\,370}
%2MASS J05352762$-$0509337}

%%author names are separated by comma (,)
%%use \and before the last author name
%%use a * along with the number separated by comma
%% for the  author for correspondence
%%\textsuperscript{number} is used for affiliation
%%\affilOne, \affilTwo etc., upto \affilTwentyfive is possible
%%Please note the first letter after \affil is capitalised in the command
%%

\author{Somnath Dutta\textsuperscript{1,2,*}, 
Koshvendra Singh\textsuperscript{3,4},
Devendra K Ojha\textsuperscript{4}, 
Saurabh Sharma\textsuperscript{3}, 
Tarak Chand\textsuperscript{3}, 
Manojit Chakraborty\textsuperscript{3} 
Kartik Gokhe\textsuperscript{3}
Joe P. Ninan\textsuperscript{4}, and 
Jyotirmoy Dey\textsuperscript{4}
% Varghese Reji\textsuperscript{1}
}
\affilOne{\textsuperscript{1}Academia Sinica Institute of Astronomy and Astrophysics, Roosevelt Rd, Taipei 106319, Taiwan, R.O.C.\\}
\affilTwo{\textsuperscript{2} Indian Centre for Space Physics, 466 Barakhola, Singabari Road, Netai Nagar, Kolkata 700099, India.\\}
\affilThree{\textsuperscript{3} Aryabhatta Research Institute of Observational Sciences (ARIES), Manora Peak, Nainital-263001, India.\\}
\affilFour{\textsuperscript{4}Department of Astronomy and Astrophysics, Tata Institute of Fundamental Research (TIFR), Mumbai-400005, India.}

%%escape two column mode for title, affiliation and abstract
%%by giving \twocolumn command as shown

\twocolumn[{

\maketitle

%%include \corres to print the corresponding author Email id
\corres{duttasomnath9@gmail.com}

%%include \msinfo for
%%manuscript information such as
%%received, revised and accepted dates
%%
\msinfo{August 2026}{October 2026}

\begin{abstract}
We present near-infrared K-band ($1.94$--$2.52~\mu$m) spectroscopy of the deeply embedded protostellar source HOPS\,370, obtained with the TIFR--ARIES Near Infrared Spectrometer (TANSPEC) on the 3.6-m Devasthal Optical Telescope (DOT). The spectrum is dominated by prominent molecular hydrogen  ($\rm H_2$) ro-vibrational emission, with multiple transitions detected from the $1$--$0$ S, $2$--$1$ S, and $1$--$0$ Q branches. Using the intrinsically linked $\rm H_2$ $1$--$0$ Q(3) and $1$--$0$ S(1) transitions, which arise from the same upper energy level, we derive a visual extinction of  $A_V = 23.24 \pm 1.25$ mag, corresponding to  $A_K = 2.50 \pm 0.13$ mag, consistent with the embedded Class~0/I nature of the source. The diagnostic line ratio $F(2$--$1~{\rm S}(1))/F(1$--$0~{\rm S}(1)) = 0.12 \pm 0.01$ is consistent with  thermal excitation and disfavors ultraviolet fluorescence as the dominant excitation mechanism. Assuming optically thin emission and local thermodynamic equilibrium, we derive an $\rm H_2$ excitation diagram that yields an excitation temperature of $2078 \pm 215$ K and a molecular hydrogen column density of $(1.93 \pm 0.93)\times10^{20}$ cm$^{-2}$. We further derive an ortho-to-para ratio of $2.45 \pm 0.21$, indicating substantial shock-driven thermal processing while remaining below the high-temperature equilibrium value of 3. Taken together, these diagnostics indicate that the molecular hydrogen emission in HOPS\,370 originates predominantly in warm, shock-heated gas most likely associated with its protostellar outflow. Our results demonstrate the diagnostic power of combined S- and Q-branch $\rm H_2$ spectroscopy for constraining the physical conditions and excitation history of deeply embedded protostellar systems.
\end{abstract}

%%insert keywords separated by 3 hyphens using \keywords{words}

}]

%%close the twocolumn escape here

%%include \doinum{number}for the DOI number in the header
%%include \volnum{number} for the volume number in the header
%%include \year{yyyy} for  year of publication in the header
%%include \pgrange{num--num} page range of article in the header
%%include \artcitid{num} for the article citation id
%%include \lp to print last page of the article
%%include \setcounter{page}{pagenum} for the exact starting page of the article

\doinum{12.3456/s78910-011-012-3}
\artcitid{\#\#\#\#}
\volnum{000}
\year{0000}
\pgrange{1--}
\setcounter{page}{1}
\lp{1}

%%%%%%%%%%%
\section{Introduction}

Stars form via the gravitational collapse of dense molecular cloud cores, passing  through deeply embedded protostellar phases before emerging onto the pre-main-sequence  track  \citep[e.g.,][]{2000prpl.conf...59A,2003ARA&A..41...57L}. During these early evolutionary stages, dense natal circumstellar envelopes strongly obscure the  central protostar at optical wavelengths  \citep[e.g.,][]{2000prpl.conf...59A,2016ApJS..224....5F}. Consequently,  near-infrared (NIR) spectroscopy serves as an indispensable tool to pierce this  obscuration and probe the immediate protostellar environment, providing vital access  to diagnostic ro-vibrational lines that trace accretion, outflows, and shock dynamics.  
%In this study, we investigate the localized feedback  mechanics and cloud-clearing dynamics of deeply embedded protostars by resolving the  precise spatial and thermal properties of their shock-excited gas.
In this study, we investigate the localized feedback and cloud-clearing processes associated with deeply embedded protostars by characterizing the spatial distribution and morphology of the shock-excited molecular gas and constraining its excitation temperature and column density.

%Molecular hydrogen ($\rm H_2$), the dominant constituent of cold molecular clouds, lacks a permanent dipole moment, rendering its cold bulk undetectable. 
 Molecular hydrogen ($\rm H_2$), the dominant constituent of cold molecular clouds, lacks a permanent dipole moment, making the cold molecular gas difficult to detect directly through H$_2$ rotational emission; instead, molecules such as CO are commonly used as tracers \citep[e.g.,][]{2013ARA&A..51..207B,2020ARA&A..58..727J}.  However, when collisional shocks or stellar radiation fields heat the gas to $T \sim 1500\text{--}3000$~K, prominent ro-vibrational transitions are excited in the NIR $K$ band \citep{2019A&A...630A..58R,2019A&A...622A.100G,1989ApJ...342..306H}, tracing the interfaces where protostellar jets interact with the surrounding molecular cloud  \citep{2006A&A...449.1077C,2024A&A...691A.134C,2001MNRAS.326..524D,2026A&A...711A.258F}. While the ubiquitous $\rm H_2$ $1\text{--}0$~S(1) line at $2.1218~\mu\text{m}$ is the standard tool for charting these outflows, line ratios involving higher-level states like $\rm H_2$ $2\text{--}1$~S(1) provide crucial diagnostics to discriminate between thermal equilibrium shocks and non-thermal ultraviolet (UV) fluorescence \citep{1990ApJ...352..625B,2015A&A...573A..82C}.

Simultaneously, the underutilized $\rm H_2$ $1\text{--}0$~Q-branch sequence ($\lambda \sim 2.40\text{--}2.50~\mu\text{m}$) provides a robust diagnostic.  Because  selected Q-branch and S-branch transitions originate from identical upper quantum states \citep[e.g.,][]{2011A&A...528A...3D}, their intrinsic line-flux ratios are determined by the corresponding transition probabilities and wavelengths, and
therefore any observed deviations can be served as a direct gauge of local, wavelength-dependent circumstellar extinction ($A_V$) \citep[e.g.,][]{1978ApJ...223..464B,2011A&A...528A...3D,2016AJ....151..173Y}. 
%Despite this diagnostic utility, the Q-branch is frequently neglected due to severe telluric water vapor degradation at the red edge of the $K$ band.
%{\bf Despite this diagnostic utility, the Q-branch is frequently neglected because its lines lie near the red edge of the $K$ band, where strong and variable telluric water-vapor absorption can significantly affect their measured fluxes.}
 Despite this diagnostic utility, measurements of the Q-branch can be challenging because its lines lie near the red edge of the $\rm K$ band, where strong and variable telluric water-vapour absorption can introduce significant systematic uncertainties in the measured fluxes \citep{1998ApJ...499..799L,2011MNRAS.411.1453G}.

In this work, we utilize the TIFR--ARIES Near Infrared Spectrometer (TANSPEC) on the 3.6-m Devasthal Optical Telescope (DOT) to analyze the deeply embedded young stellar object (YSO) HOPS 370 (also known as, 2MASS J05352762$-$0509337). 
HOPS 370 is a luminous ($L_{\text{bol}} \approx 315-361\ L_\odot$, K = 10.777 mag), intermediate-mass Class\,I protostar located within the Orion A molecular cloud \citep{2016ApJS..224....5F}. With a bolometric temperature of $T_{\text{bol}} \approx 71.5\text{ K}$, it occupies the critical evolutionary transition boundary separating deeply embedded Class 0 objects from more evolved circumstellar disk systems \citep{2020ApJ...890..130T}.  High-resolution interferometric observations reveal a massive, rotationally supported protostellar dust disk ($M_{\text{disk}} \approx 114.6\ M_\oplus$) with a characteristic radius of $R_{\text{disk}} \approx 112.8\text{ au}$ \citep{2020ApJ...890..130T}.  Furthermore, it drives a powerful, highly collimated bipolar molecular outflow that carves distinct opacity cavities into its ambient envelope \citep{2024ApJ...966...41F,2025ApJ...983..110T}, making HOPS 370 an excellent target for studying localized shock chemistry and stellar feedback. 

The paper is organized as follows. In Section 2,
%~\ref{sec:ObservationsandDataReduction}, 
we present the new $K$-band spectroscopic observations and the corresponding data  reduction procedures. Section 3 %~\ref{sec:AnalysesAndResults} 
describes the data 
analysis and the estimation of various physical parameters derived from the  observed spectra, which are used to dynamically constrain the physical conditions  of the inner envelope. In Section 4, %~\ref{sec:discussions}, 
we place our findings  into a broader astrophysical context through a detailed discussion. Finally,  we summarize our main conclusions in Section 5. %~\ref{sec:conclusions}.

\begin{figure}
    \centering
    \includegraphics[width=1\linewidth]{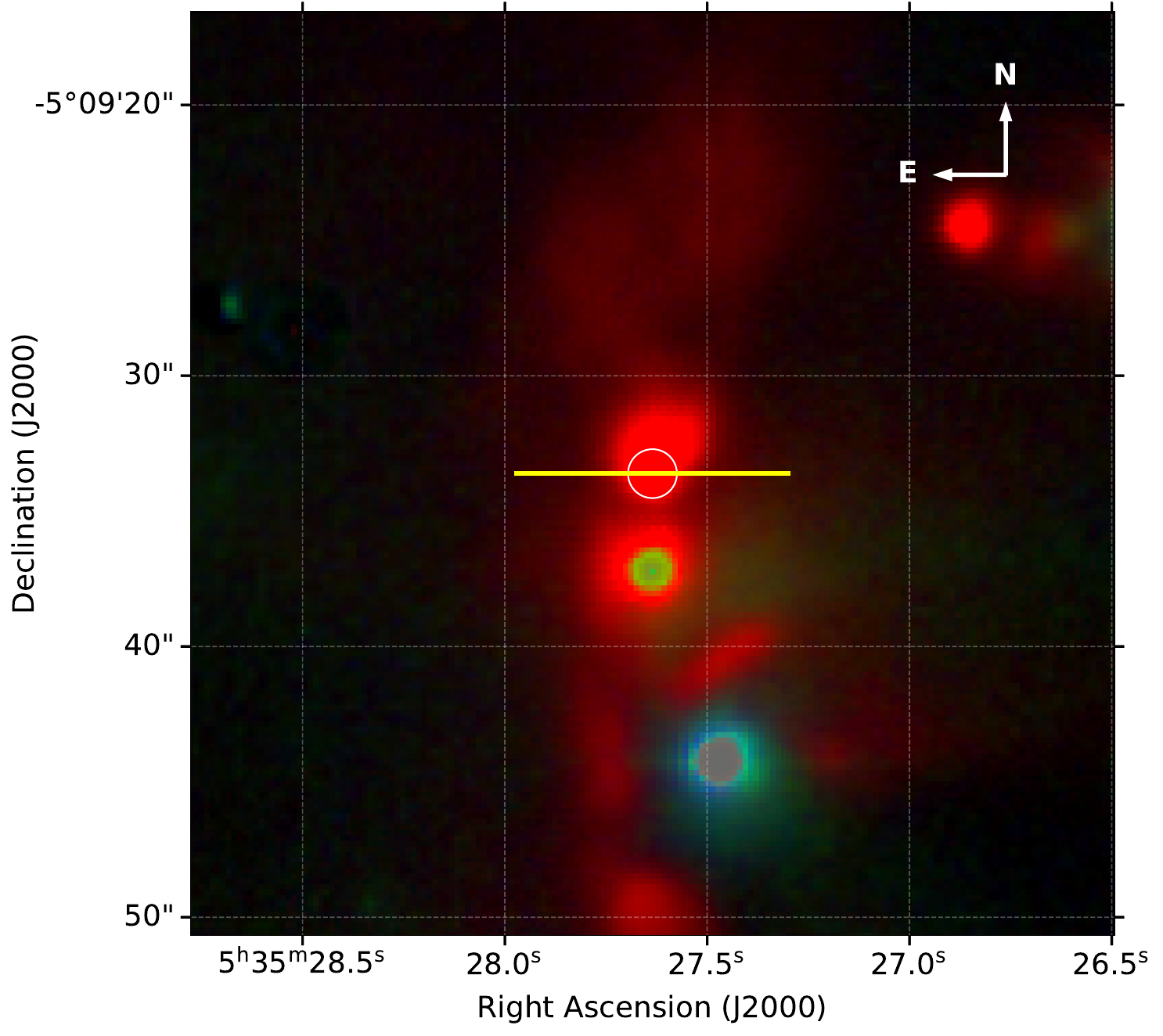}
\caption{Three-color composite image of the target field HOPS\,370 generated from UKIDSS data. Red, green, and blue channels correspond to the $K$, $J$, and $Y$ bands, respectively. The target source position is highlighted at the center with white circle. The yellow bar denotes orientation of slit position for spectroscopic observations.}
    \label{fig:ukidss_rgb_composite}
\end{figure}

\section{Observations and Data Reduction}\label{sec:ObservationsandDataReduction}

\subsection{Observations}

Near-infrared spectroscopic observations of HOPS\,370 were carried out at continuum peak  %Taken from other place: $\alpha = 05^{\mathrm{h}}\,35^{\mathrm{m}}\,27.61^{\mathrm{s}}$, $\delta = -05^{\circ}\,09'\,34.5''$ 
%5:35:27.6352 -5:09:33.616 taken from obseved image+UKIDSS
$\alpha = 05^{\mathrm{h}}\,35^{\mathrm{m}}\,27.635^{\mathrm{s}}$, 
$\delta = -05^{\circ}\,09'\,33.616''$ \citep{2020ApJ...890..130T} (Figure \ref{fig:ukidss_rgb_composite})  using the TIFR--ARIES Near Infrared Spectrometer (TANSPEC) mounted on the 3.6-m Devasthal Optical Telescope (DOT), operated by the Aryabhatta Research Institute of Observational Sciences (ARIES), Nainital, India. TANSPEC is a cross-dispersed spectrograph providing simultaneous wavelength coverage from approximately 0.55 to 2.5~$\mu$m in a single exposure \citep{2022PASP..134h5002S}.

%The target was observed on  \texttt{date} = \text{2026-02-17},  \texttt{time} = \text{08:37:43.536} \text{(UT)} 
The target was observed on 17 February 2026 at 08:37:43 UT using the cross-dispersed mode with a slit width of $0.5\,\text{arcsec}$, yielding a spectral  resolving power of $R = \lambda/\Delta\lambda \sim 2500$. The spatial profile across the cross-dispersion direction yielded a seeing of $2.3''$ during the observations.
%The observing conditions were photometric with a typical seeing of $\sim$1.0 arcsec. 
%Calibration frames including flat-field exposures and wavelength calibration lamps, standard star  were obtained as part of the standard TANSPEC observing sequence.
Calibration frames, including flat-field exposures and wavelength calibration lamps, as well as standard-star observations, were obtained.
\subsection{Data Reduction}

The spectroscopic data were reduced using the standard \texttt{pyTANSPEC} (v1.0) pipeline designed for cross-dispersed spectra \citep{2026JApA...47...40R}.  The reduction workflow involved standard procedures, including dark subtraction, flat-field correction, cosmic-ray rejection, and background sky subtraction. One-dimensional spectral extraction was performed, followed by wavelength calibration using an arc lamp assembly and subsequent telluric correction using observations of a standard star.

Although the full TANSPEC setup covers a wide wavelength range from approximately $0.55$ to $2.5\,\mu\text{m}$, this study focuses exclusively on the $K$-band region spanning $1.94$--$2.52\,\mu\text{m}$. The extracted $K$-band spectrum was normalized to the local continuum to facilitate line identification and detailed analysis of the molecular hydrogen ($\text{H}_2$) emission features.

\subsection{Archival Data}

Complementary near-infrared $Y$-, $J$-, and $K$-band photometric observations of the target region were retrieved from the UKIRT Infrared Deep Sky Survey \citep[UKIDSS;][]{2007MNRAS.379.1599L}. These data were acquired using the Wide Field Camera \citep[WFCAM;][]{2007A&A...467..777C} mounted on the 3.8-m United Kingdom Infrared Telescope (UKIRT). The high-resolution archival images provide spatial and photometric context, enabling a detailed examination of the surrounding stellar distribution and the identification of potential stellar or substellar companions near the target. The RGB image constructed from the $Y$, $J$, and $K$ bands is shown in Figure \ref{fig:ukidss_rgb_composite}. To contextualize the multi-wavelength analysis, the precise pointing center of our TANSPEC spectroscopic observations and the relative orientation of the $0.5''$ slit are overlaid on the color composite.

\section{Analyses and Results}\label{sec:AnalysesAndResults}
\begin{figure*}
\centering
\includegraphics[width=0.95\textwidth]{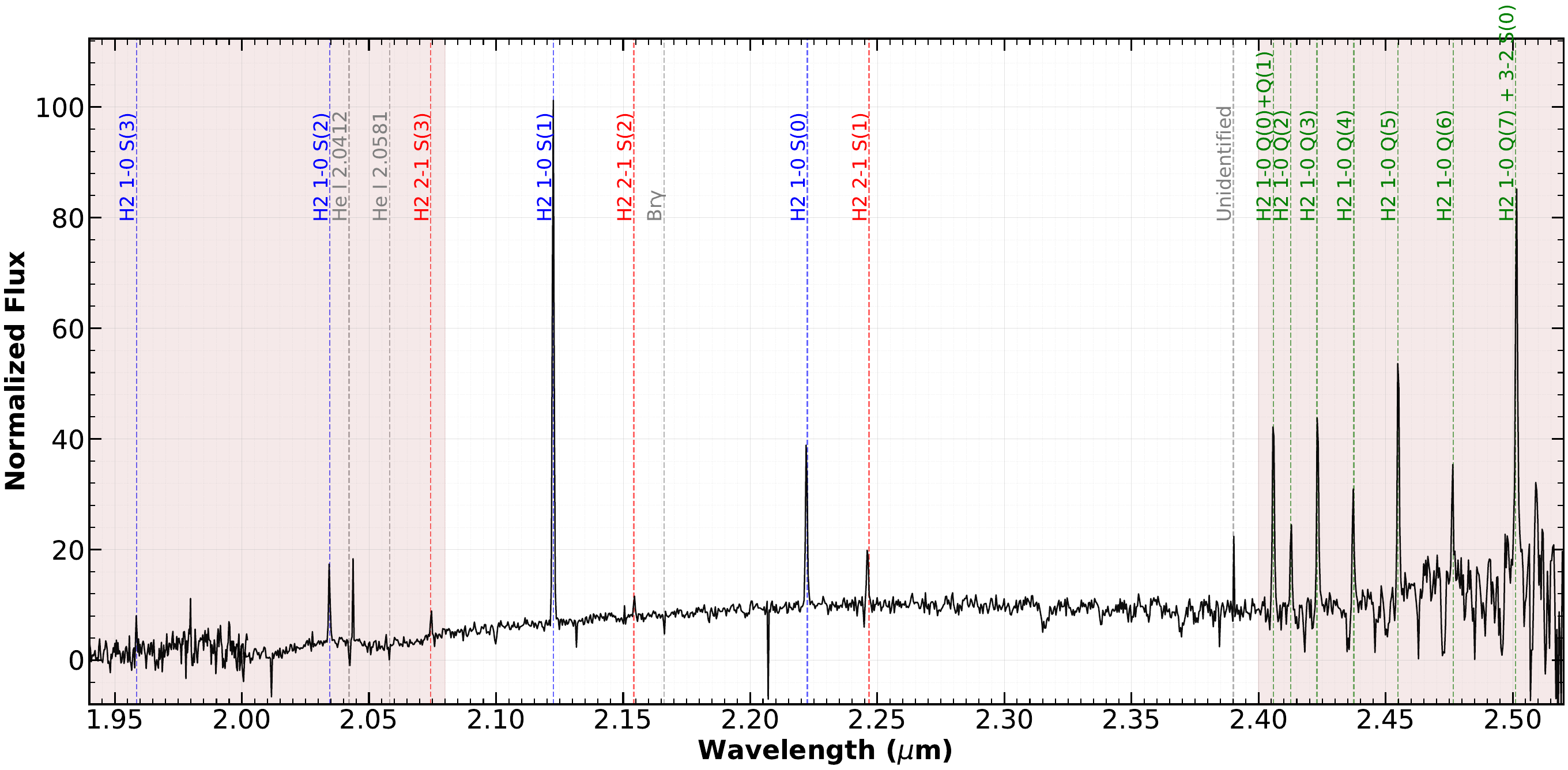}
\caption{
Full K-band  continuum-subtracted spectrum of HOPS\,370 obtained with TANSPEC.
The principal H$_2$ ro-vibrational transitions identified in the spectrum are marked.
The spectrum is dominated by molecular hydrogen emission, including both S-branch
and Q-branch transitions. The shaded regions indicate the  wavelength ranges likely affected by telluric atmospheric absorption.
}
\label{fig:kspec}
\end{figure*}
% \subsection{K-band Spectrum}

\subsection{K-band Spectrum}

The reduced K-band spectrum of HOPS\,370 is presented in Figure~\ref{fig:kspec}. The spectrum is dominated by numerous molecular hydrogen emission features superimposed on a relatively weak continuum. Several H$_2$ ro-vibrational transitions are clearly detected, including lines from the 1--0 and 2--1 S-branch series, as well as a sequence of H$_2$ Q-branch transitions between 2.4 and 2.5~$\mu$m.

%To verify the reality of the detected features, all candidate emission lines were visually inspected in the corresponding two-dimensional spectra. Residual telluric features in 2D spectrum typically appear as vertical strips with equal brightness, whereas genuine source emission is confined to the stellar position and appears as a localized bright feature. Based on this criterion, all identified H$_2$ lines were confirmed to be astrophysical in origin rather than artifacts of imperfect telluric features. {\bf This verification is particularly important for the Q-branch transitions at the red end of the $K$-band, where strong atmospheric absorption may complicate line identification.} The excellent agreement between the observed line positions, the expected sequence of H$_2$ Q-branch transitions, and their spatial localization in the two-dimensional spectra provides strong evidence that these features are genuine molecular hydrogen emission lines.

 To verify the reality of the detected features, all candidate emission lines were visually inspected in the corresponding two-dimensional spectra and cross-checked against their expected wavelengths. Residual telluric features can affect near-infrared spectra and may complicate the identification and measurement of weak astronomical emission lines \citep[e.g.,][]{2003PASP..115..389V}. In our two-dimensional spectra, residual telluric features generally extend along the spatial direction of the slit, whereas genuine source emission is localized at the source position. Based on these criteria, the identified $\rm H_2$ lines are considered astrophysical in origin rather than artifacts of imperfect telluric correction. This verification is particularly important for the Q-branch transitions at the red end of the $\rm K$ band, where atmospheric transmission is poor and variable beyond 2.4 $\mu$m, making reliable measurements of these lines more challenging \citep{2011MNRAS.411.1453G}.

%The  line at 2.1218~$\mu$m H2 (1--0) S(1) is the brightest one in the spectra. 

Line profiles were measured by mapping the local adaptive spline continua, with the final relative and absolute line fluxes calculated by directly integrating the residual emission. 
%Relative line fluxes were converted to absolute flux units using the observed $K$-band magnitude of the source. 
All identified spectral features were visually inspected and cross-checked against published near-infrared line catalogs of young stellar objects and molecular outflows.  The detected transitions are listed in Table~\ref{tab:h2lines}.

The extracted spectrum was flux calibrated using the $K$-band magnitude of HOPS\,370 from the Visible and Infrared Survey Telescope for Astronomy \citep[VISTA;][]{2019A&A...622A.149G} survey, converting the relative spectral fluxes to absolute units through the standard $K$-band zero-point calibration. No explicit slit-loss correction was applied; therefore, the absolute line fluxes may be subject to additional systematic uncertainties associated with slit throughput and source variability. Extinction corrections were not applied during the calibration process and were instead incorporated later through the H$_2$ line-ratio analysis.

The spectrum is characterized by strong molecular emission accompanied by comparatively weak hydrogen recombination lines and \Hei\ emission. Notably, the \Hei\ $\lambda$2.0412 and $\lambda$2.0581 transitions show tentative asymmetric profiles that may be consistent with P-Cygni-type structure.
%
%appear to exhibit likely P-Cygni profiles, while the Br$\gamma$ line at 2.1661~$\mu$m is weak and predominantly seen in absorption. 
%
The Br$\gamma$ line at 2.1661~$\mu$m appears weak and may be present in absorption; however, given the moderate spectral resolution and signal-to-noise ratio of the data, we do not attempt a detailed interpretation of this feature. The overall $K$-band spectral morphology suggests that the emission is dominated by warm molecular gas rather than active magnetospheric accretion. The detection of numerous H$_2$ transitions provides a valuable opportunity to investigate the excitation conditions and physical origin of the molecular emission.

\begin{table*}[htbp]
  \centering
  \caption{Line Measurement Parameters, Transition Configurations, and Derived Absolute Fluxes}
  \label{tab:h2lines}
  \begin{tabular}{lccccccc}
    \toprule
    Transition & $\lambda_{\text{rest}}$  & $A_{ul}$ & $g_u$ & $E_u/k$  & Absolute Flux & State / Note \\
               &     ($\mu$\text{m})     &               ($\times 10^{-7}$ s)          &     &              (\text{K})           & (\text{erg s}$^{-1}$ \text{cm}$^{-2}$) & \\
    \midrule
    && Atomic&&&\\
    \midrule
    Br$\gamma$     & 2.1661   & --                     & --    & --       & -- & Absorption \\
    He I 2.0412    & 2.0422   & --                     & --    & --       & -- & P-Cygni \\
    He I 2.0581    & 2.0581   & --                     & --    & --       & -- & P-Cygni \\
    \midrule
    && Molecular&&&\\ 
    \midrule
    H$_2$ 1-0 S(0) & 2.2235   & 2.53     & 5     & 6471     & $(5.99 \pm 0.18) \times 10^{-14}$ & para \\
    H$_2$ 1-0 S(1) & 2.1218   & 3.47    & 21    & 6956     & $(1.86 \pm 0.02) \times 10^{-13}$ & ortho \\
    H$_2$ 1-0 S(2) & 2.0338   & 3.98   & 9     & 7584     & $(2.28 \pm 0.19) \times 10^{-14}$ & para \\
    H$_2$ 1-0 S(3) & 1.9586   & 4.21   & 33    & 8365     & $(8.13 \pm 0.34) \times 10^{-14}$ & ortho \\
    &&&&&\\
    H$_2$ 2-1 S(1) & 2.2477   & 4.98    & 21    & 12550    & $(2.23 \pm 0.18) \times 10^{-14}$ & ortho \\
    H$_2$ 2-1 S(2) & 2.1542   & 5.60    & 9     & 13150    & $(6.54 \pm 1.84) \times 10^{-15}$ & para \\
    H$_2$ 2-1 S(3) & 2.0735   & 5.77    & 33    & 13890    & $(7.42 \pm 1.89) \times 10^{-15}$ & ortho \\
    &&&&&\\
    H$_2$ 1-0 Q(0)+Q(1) & 2.4066   &  --   &  --     &  --     & $(7.62 \pm 0.19) \times 10^{-14}$ & blended \\
    H$_2$ 1-0 Q(2) & 2.4134   & 3.03    & 5     & 6471     & $(7.15 \pm 0.19) \times 10^{-14}$ & para \\
    H$_2$ 1-0 Q(3) & 2.4237   & 2.78    & 21    & 6956     & $(8.16 \pm 0.19) \times 10^{-14}$ & ortho \\
    H$_2$ 1-0 Q(4) & 2.4375   & 2.65    & 9     & 7566     & $(4.56 \pm 0.19) \times 10^{-14}$ & para \\
    H$_2$ 1-0 Q(5) & 2.4548   & 2.55    & 33    & 8365     & $(1.02 \pm 0.02) \times 10^{-13}$ & ortho \\
    H$_2$ 1-0 Q(6) & 2.4756   & 2.45    & 13    & 9286     & $(4.64 \pm 0.19) \times 10^{-14}$ & para \\
    H$_2$ 1-0 Q(7) + 3-2 S(0) & 2.5001   &  --   &  --    &  --    & $(1.76 \pm 0.02) \times 10^{-13}$ & blended \\
    \bottomrule
  \end{tabular}
  \tablenotes{
The line identifications, Einstein $A$ coefficients, and energy levels are adopted from the \texttt{NIST Standard Atomic Spectra Database}\footnote{https://www.nist.gov/pml/atomic-spectra-database}\citep{NIST_ASD} and the \texttt{Gemini Important $\rm H_2$ Line List}\footnote{https://www.gemini.edu/observing/resources/near-ir-resources/spectroscopy/important-h2-lines}.} 
\end{table*}

\subsection{Detected H$_2$ Emission Lines}

As summarized in Table~\ref{tab:h2lines}, the observed transitions encompass an extensive suite of both S-branch and Q-branch lines originating from vibrationally excited molecular states. The simultaneous detection of these transitions allows for a detailed investigation into the excitation temperature, visual extinction ($A_V$), and underlying excitation mechanisms of the emitting gas.

The dominant feature in the spectrum is the H$_2$ 1--0 S(1) transition at 2.1218~$\mu$m, a classic tracer commonly associated with shocked molecular gas in protostellar outflows  \citep[e.g.,][]{2001ARA&A..39..403R}. Beyond this primary feature, several other fundamental rovibrational transitions are detected, including lower S-branch lines from the 1--0 vibrational state, higher-excitation 2--1 S-branch lines, and an array of 1--0 Q-branch lines longward of 2.40~$\mu$m. This broad collection of transitions spans a wide range of upper-state energies ($E_u/k$), providing the empirical basis needed to construct an H$_2$ excitation diagram and distinguish between thermal shock excitation and non-thermal ultraviolet fluorescence.

\subsection{Extinction from H$_2$ Q- and S-Branch Transitions}

A powerful advantage of detecting both Q-branch and S-branch transitions is that several of these lines originate from the same upper rovibrational level.  Under optically thin conditions, which are generally appropriate for the weak electric-quadrupole $\rm H_2$ rovibrational transitions at the column densities characteristic of shock-excited gas \citep[e.g.,][]{2025ApJ...995..199N,2025ApJ...985..225L}, their intrinsic line ratios are determined solely by the transition probabilities and wavelengths, making them largely insensitive to uncertainties in excitation conditions or localized gas temperatures. In particular, the H$_2$ 1--0 Q(3) and 1--0 S(1) transitions share the upper level $(v,J)=(1,3)$. Their intrinsic flux ratio is given by:

\begin{equation}
\left(\frac{F_{Q(3)}}{F_{S(1)}}\right)_{\rm int} = \frac{A_{Q(3)}\lambda_{S(1)}}{A_{S(1)}\lambda_{Q(3)}},
\end{equation}

where $A_{ul}$ denotes the Einstein transition probability from the upper level ($u$) to the lower level ($l$) and $\lambda$ is the rest wavelength of the transition. Consequently, any deviation between the observed line ratio and this intrinsic quantum mechanical value can be attributed primarily to wavelength-dependent differential extinction. 

We derived the visual extinction from the H$_2$ 1--0 Q(3)/S(1) line ratio using a near-infrared power-law extinction curve \citep{2005ApJ...619..931I}. The extinction coefficient (k($\lambda$)), defined as the ratio of the extinction at wavelength ($\lambda$) to the visual extinction (A$_V$), is given by $k(\lambda) = A_\lambda / A_V$ = 
$0.112
\left(\frac{\lambda}{2.148\,\mu{\rm m}}\right)^{-1.61},
$
corresponding to an extinction-law normalization of $A_K/A_V = 0.112$ and a power-law index $\alpha = 1.61$.

% We derived the visual extinction from the H$_2$ 1--0 Q(3)/S(1) line ratio using a near-infrared power-law extinction curve \citep{2005ApJ...619..931I},
% 0.112
% $\left(\frac{\lambda}{2.148,\mu{\rm m}}\right)^{-1.61},
% ]$
% corresponding to an extinction-law normalization of $A_K/A_V = 0.112$ and a power-law index $\alpha = 1.61$.

% Using a standard near-infrared power-law extinction law where $k(\lambda) = A_\lambda / A_V$  \citep[e.g.,][]{2005ApJ...619..931I}, 
The visual extinction $A_V$ toward the embedded protostar can be estimated as:

\begin{equation}
A_V = \frac{2.5}{k(\lambda_{S1}) - k(\lambda_{Q3})} \log_{10} \left[ \frac{(F_{Q(3)}/F_{S(1)})_{\rm int}}{(F_{Q(3)}/F_{S(1)})_{\rm obs}} \right],
\end{equation}

where $(F_{Q(3)}/F_{S(1)})_{\rm obs}$ is the observed absolute flux ratio.

From this relation, we derived a visual extinction of $A_V = 23.24 \pm 1.25$\,mag, corresponding to a $K$-band extinction of $A_K = 2.50 \pm 0.13$\,mag. 
%Because the $\mathrm{Q}$-branch region ($2.4\text{--}2.5\,\mu\text{m}$) is prone to telluric absorption, a $10\text{--}20\%$ telluric deficit in $F_{\mathrm{Q}(3)}$ introduces a severe shift of $\Delta A_V \approx +4.93 \text{ to } +9.86$\,mag ($\Delta A_K \approx +0.53 \text{ to } +1.06$\,mag), underscoring the strong leverage differential flux uncertainties exert on this diagnostic.
%
This high level of extinction is characteristic of deeply embedded Class~0/I protostars and is consistent with the range ($A_V \approx 10\text{--}30$\,mag) measured in dense star-forming environments observed with \textit{JWST} \citep[e.g.,][]{2026MNRAS.548ag572L}, as well as in individual protostars studied by the \textit{JWST} Observations of Young protoStars (JOYS) program \citep[e.g.,][]{2025A&A...699A.361V}. Comparable extinction values have also been reported for low-mass protostars such as IRAS~16253$-$2429 \citep{2024ApJ...962L..16N} and HOPS~315 \citep{2025ApJ...991...45D}, both of which drive highly collimated jets and molecular outflows while remaining embedded within their natal envelopes. The derived extinction toward HOPS\,370 supports its classification as an actively accreting protostellar system still deeply embedded in its parent molecular cloud.

%So a systematic uncertainty of only: ±10% in Q(3) produces roughly ±5 mag in A_V = ±20% in Q(3) produces roughly −9/+12 mag in A_V.
The extinction estimate derived from the H$_2$ 1--0 Q(3)/S(1) line ratio should be interpreted with some caution, as the Q-branch transitions lie in a wavelength region affected by strong telluric absorption. Residual telluric-correction uncertainties at the 10--20\% level could introduce systematic uncertainties of several magnitudes in the derived extinction. For the present dataset, a 20\% uncertainty in the Q(3) flux would result in $A_V$ values ranging from approximately 14 to 35 mag.

\begin{figure*}
    \centering
    \includegraphics[width=0.66\textwidth]{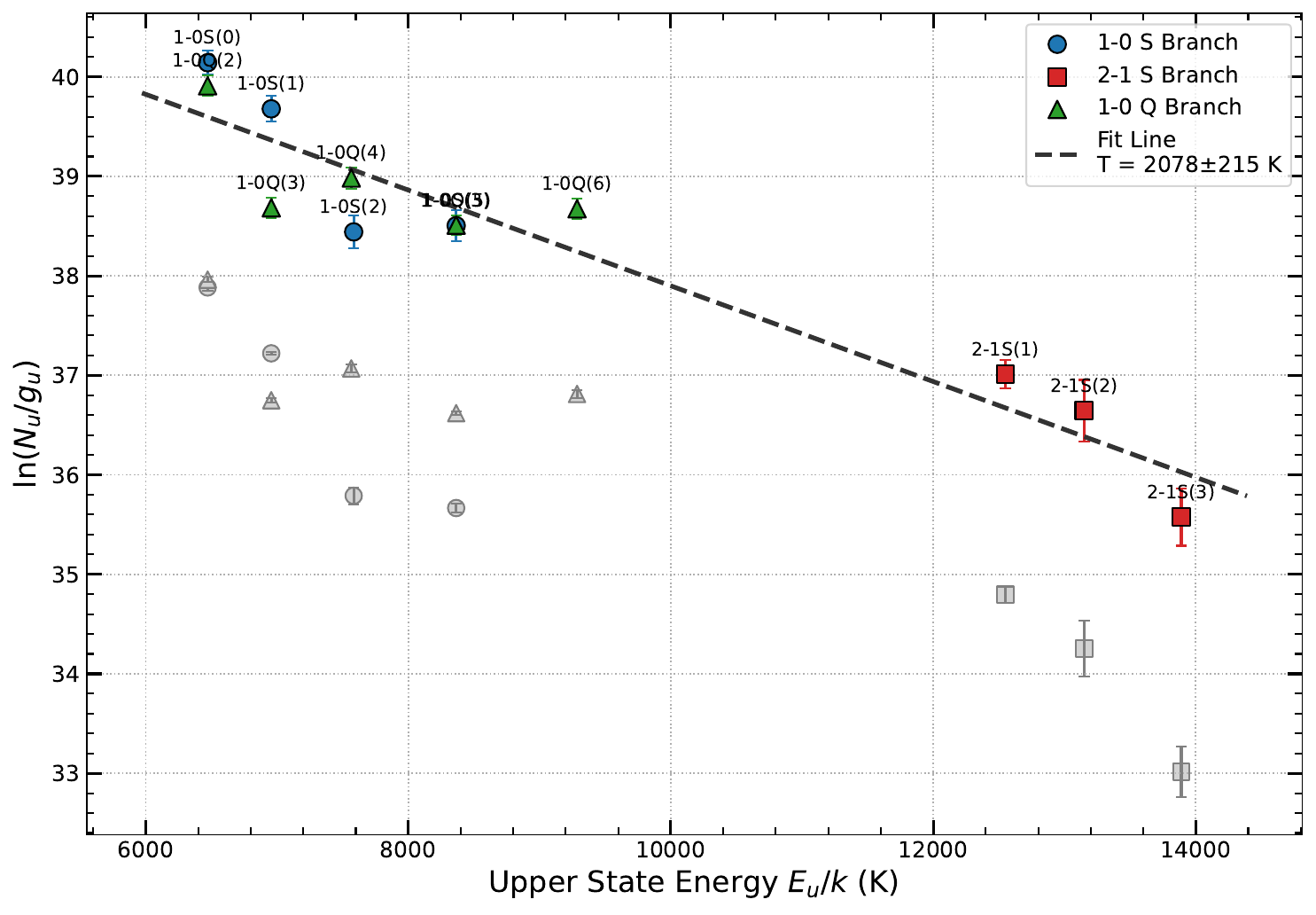}
    \caption{Molecular hydrogen ($\rm H_2$) rovibrational excitation diagram derived from $K$-band line measurements. The natural logarithm of the upper-state column density normalized by its statistical weight, $\ln(N_u/g_u)$, is shown as a function of the upper-state energy $E_u/k$ in Kelvin. Gray markers represent the baseline observed data points, while colored markers denote data corrected for interstellar extinction. Data points are grouped by transitions: 1-0 S branch (blue circles), 2-1 S branch (green squares), and 1-0 Q branch (purple triangles). The dashed crimson line illustrates an unweighted ordinary least-squares linear fit applied to the dereddened population distribution, yielding an excitation temperature of $T_{\text{ex}} = \text{2078} \pm \text{215}$~K and column density $\rm N_{H_2} = (1.927 \pm 0.93) \times 10^{20} cm^{-2}$. Vertical error bars reflect the combined quadrature sum of relative flux measurement uncertainties and propagated errors from the extinction correction.  The fit has an RMS residual scatter of $\sigma_{\text{scatter}} \approx 0.41$, with a reduced chi-squared of $\chi^2_{\nu} \approx 12.4$.}
    \label{fig:h2_excitation}
    % \caption{H2 Rotational Diagram}
    % \label{fig:H2_excitation_temperature_diagram}
\end{figure*}

\subsection{Shock versus Fluorescent Excitation}

The excitation mechanism of the molecular hydrogen emission was investigated using the ratio of the vibrationally excited H$_2$ lines \citep{1987ApJ...322..412B, 2002MNRAS.331..154R}.  A commonly used diagnostic is the H$_2$ 2-1 S(1) to 1-0 S(1) line ratio ($R$), defined as

\begin{equation}
R = \frac{F(2\mbox{--}1\,\text{S}(1))}{F(1\mbox{--}0\,\text{S}(1))},
\end{equation}

which helps examine the excitation mechanism. Because shock excitation and UV fluorescence populate the rovibrational levels through different physical processes and therefore produce distinct relative line strengths. Specifically, because the $v=2$ level requires more energy than the $v=1$ level, its relative thermal population is low, yielding a small value of $R$ ($\lesssim 0.2$) characteristic of thermal shock excitation. Conversely, UV fluorescence pumps molecules into excited electronic states, with subsequent radiative cascades heavily overpopulating higher vibrational levels ($v=2, 3$, etc.) relative to a thermal distribution. This results in a significantly higher value of $R$ (often approaching $\sim 0.5\text{--}0.6$), making this ratio a powerful tool for distinguishing between thermal and non-thermal excitation regimes.
Using the measured line fluxes listed in Table~\ref{tab:h2lines}, we obtain $R = 0.12 \pm 0.01$. This value is significantly lower than the typical ratios expected for ultraviolet fluorescent excitation ($R \sim 0.5$--$0.6$) and is consistent with thermal excitation in shocked molecular gas ($R \lesssim 0.2$) \citep[e.g.,][]{2002MNRAS.331..154R,2020MNRAS.492..294G}.

The vibrational excitation temperature was estimated directly from the column densities of the $v=1,\ J=3$ and $v=2,\ J=3$ levels, corresponding to the H$_2$ 1--0 S(1) and 2--1 S(1) transitions, respectively. Assuming a Boltzmann distribution of level populations, the vibrational temperature is given by

\begin{equation}
\frac{N(v=2,J=3)}
     {N(v=1,J=3)}
=
\exp\left[
-\frac{E_{2,3}-E_{1,3}}
       {T_{\rm vib}}
\right],
\end{equation}

 where $N(v,J)$ denotes the upper-level column density,  $E_{2,3}/k = 12550$~K and $E_{1,3}/k = 6956$~K. Rearranging yields

\begin{equation}
T_{\rm vib}
=
\frac{E_{2,3}-E_{1,3}}
{\ln\left[N(v=1,J=3)/N(v=2,J=3)\right]}.
\end{equation}

 Using the measured line fluxes listed in Table~\ref{tab:h2lines}, we derive the column densities of the upper rovibrational levels, $N(v,J)$, assuming optically thin emission, following \citep[e.g.,][]{2025A&A...695A.145V,2025ApJ...985..225L,2025ApJ...991...45D}:
\begin{equation}\label{equ:columndensity}
N(v,J) 
=
\frac{4\pi F_{ul}\lambda}
     {A_{ul}hc},
\end{equation}
where $u$ and $l$ denote the upper and lower rovibrational energy levels of the transition, respectively, $F_{ul}$ is the measured line flux, $\lambda$ is the rest wavelength, and $A_{ul}$ is the Einstein $A$-coefficient. We derive a vibrational temperature of $T_{\rm vib} = 2305 \pm 79$~K. The low value of the diagnostic line ratio ($R = 0.12$), together with the derived vibrational temperature, indicates that the H$_2$ emission is predominantly thermally excited. Such temperatures are characteristic of warm molecular gas heated by shocks most likely associated with protostellar jets and outflows. The dominance of strong H$_2$ emission lines and the relative weakness of Br$\gamma$ further support a shock-driven origin for the observed near-infrared spectrum.
The measured line ratio favors thermal excitation within a dense, shock-heated environment, where the molecular gas is expected to approach Local Thermodynamic Equilibrium (LTE). Under these conditions, the rotational and vibrational level populations become thermalized, leading to a common excitation temperature. Motivated by this result, we construct a unified  $\rm H_2$ rovibrational excitation diagram using all detected extinction-corrected 1--0~S, 2--1~S, and 1--0~Q transitions.

%%%%%%%%%%%%%%%%%

\subsection{Excitation Diagram}\label{sec:GlobalExcitationDiagram}

% The excitation conditions of the molecular gas were investigated using the observed H$_2$ line intensities. Assuming optically thin emission, the column density of the upper level is given by {\bf \citep[e.g.,][]{2025A&A...695A.145V,2025ApJ...985..225L,2025ApJ...991...45D}}

% \begin{equation}
% N_u
% =
% \frac{4\pi F_{ul}\lambda}
%      {A_{ul}hc},
% \end{equation}

% %where $F_{ul}$ is the measured line flux.
% {\bf where $u$ and $l$ denote the upper and lower rovibrational energy levels of the transition, respectively, $F_{ul}$ is the measured line flux, $\lambda$ is the rest wavelength, and $A_{ul}$ is the Einstein $A$-coefficient.}

% Under local thermodynamic equilibrium (LTE), the population distribution follows the Boltzmann relation

% \begin{equation}
% \ln\left(\frac{N_u}{g_u}\right)
% =
% -\frac{E_u}{kT_{\rm ex}}
% +
% C,
% \end{equation}

% where $g_u$ is the statistical weight of the upper level, $E_u$ is the excitation energy, and $T_{\rm ex}$ is the excitation temperature.

% An excitation diagram was constructed by plotting $\ln\left(\frac{N_u}{g_u}\right)$ against $\frac{E_u}{k}$, as shown Figure \ref{fig:h2_excitation}. For thermalized gas, the data points are expected to lie along a straight line whose slope yields, ${\rm slope}$ = $-\frac{1}{T_{\rm ex}}$. The derived excitation temperature provides insight into the heating mechanism operating within the molecular gas.

 The excitation conditions of the molecular gas were investigated using the observed H$_2$ line intensities.   The column density $N_u$ (or $N(v,J)$) is estimated using Equation~\ref{equ:columndensity}. Under local thermodynamic equilibrium (LTE), the population distribution follows the Boltzmann relation

\begin{equation}
\ln\left(\frac{N_u}{g_u}\right)
=
-\frac{1}{T_{\rm ex}} \left(\frac{E_u}{k}\right)
+
C,
\end{equation}

where $g_u$ is the statistical weight of the upper level, $E_u/k$ is the upper-level energy expressed in units of temperature, $k$ is the Boltzmann constant, and $T_{\rm ex}$ is the excitation temperature. The intercept constant $C$ is explicitly defined as

\begin{equation}
C = \ln\left(\frac{N_{\text{tot}}}{Z(T_{\text{ex}})}\right),
\end{equation}

where $N_{\text{tot}}$ is the total column density of molecular hydrogen summed over all rovibrational states, and $Z(T_{\text{ex}})$ is the total internal partition function of H$_2$ evaluated at the derived excitation temperature. Thus, the total column density $\rm N_\text{tot}$ (or $\rm N_{H_2}$)  is determined from the intercept $C$ through $\rm N_{\text{tot}} = \exp(C) \times Z(T_{\text{ex}})$.

An excitation diagram was constructed by plotting $\ln\left(\frac{N_u}{g_u}\right)$ against $E_u/k$, as shown in Figure \ref{fig:h2_excitation}. For thermalized gas, the data points lie along a straight line whose slope yields $\text{slope} = -1/T_{\text{ex}}$, directly providing the excitation temperature. Furthermore, the total column density  $\rm N_{H_2}$ is determined from the intercept $C$ of the best-fit line, which accounts for the molecular population distributed across all unobserved states.

We present a unified $\rm H_2$  rovibrational diagram (Boltzmann plot) across all detected transitions. The observed fluxes were corrected for extinction. By plotting  the column densities over a wide baseline of upper-state energies  ($E_u/k_B \sim 6400\text{--}14000$~K), the global linear regression filters out  local line variations, yielding an excitation temperature of $\rm T_{ex} = 2078 \pm 215$~K and column density  $\rm N_{H_2} = (1.927 \pm 0.93) \times 10^{20} cm^{-2}$ from the extinction corrected data.  To evaluate the validity of the single-temperature LTE assumption, we examine the residual scatter of the data points around the linear fit. Following the methodology of \citet{2018MNRAS.481.5370M,2026arXiv260922846P}, the global fit exhibits an RMS residual scatter of $\sigma_{\text{scatter}} \approx 0.41$ (with a reduced chi-squared of $\chi^2_{\nu} \approx 12.4$). While smaller scatter is broadly consistent with a dominant single-temperature LTE component, larger dispersions can reflect temperature gradients, multiple excitation components, or unresolved substructure \citep[e.g.,][]{2018MNRAS.481.5370M}.

Individual transition branch fittings of 1--0 S, 1--0 Q and 2--1 S were rejected in favor of a unified global  rovibrational diagram for three primary reasons. First, isolating single branches  (such as the $1\text{--}0$~S transitions) restricts the data to a narrow upper-state  energy baseline ($\Delta E_u/k_B \sim 2000 - 3000$~K), leaving the linear regression highly  vulnerable to minor observational flux scatter. Combining all detected transitions  expands this baseline to a robust range of $\Delta E_u/k_B \sim 7500$~K,  which stabilizes the fit against local noise. Second, individual fits fail to  expose systematic, wavelength-dependent anomalies, such as telluric degradation or differential extinction affecting the $1\text{--}0$~Q branch at $\lambda > 2.4~\mu\text{m}$.  Finally, fitting individual branches with sparse line counts (e.g., the $2\text{--}1$~S branch)  suffers from severely limited degrees of freedom. 

% \subsection{Ortho-to-Para Ratio}\label{sec:excitation}

\subsection{Ortho-to-Para Ratio}\label{sec:opr}

The ortho-to-para ratio (OPR) of molecular hydrogen is a valuable diagnostic of the thermal and chemical history of shocked molecular gas. Under local thermodynamic equilibrium (LTE) at temperatures above $\sim 2000$~K, the OPR approaches the statistical value of 3, whereas lower values can indicate incomplete thermalization, recent H$_2$ formation on dust grains, or gas that has not yet fully equilibrated following a shock event. Measuring the OPR therefore provides an independent probe of the excitation conditions and thermal history of protostellar outflows.

\begin{figure*}
    \centering
    \includegraphics[width=0.7\linewidth]{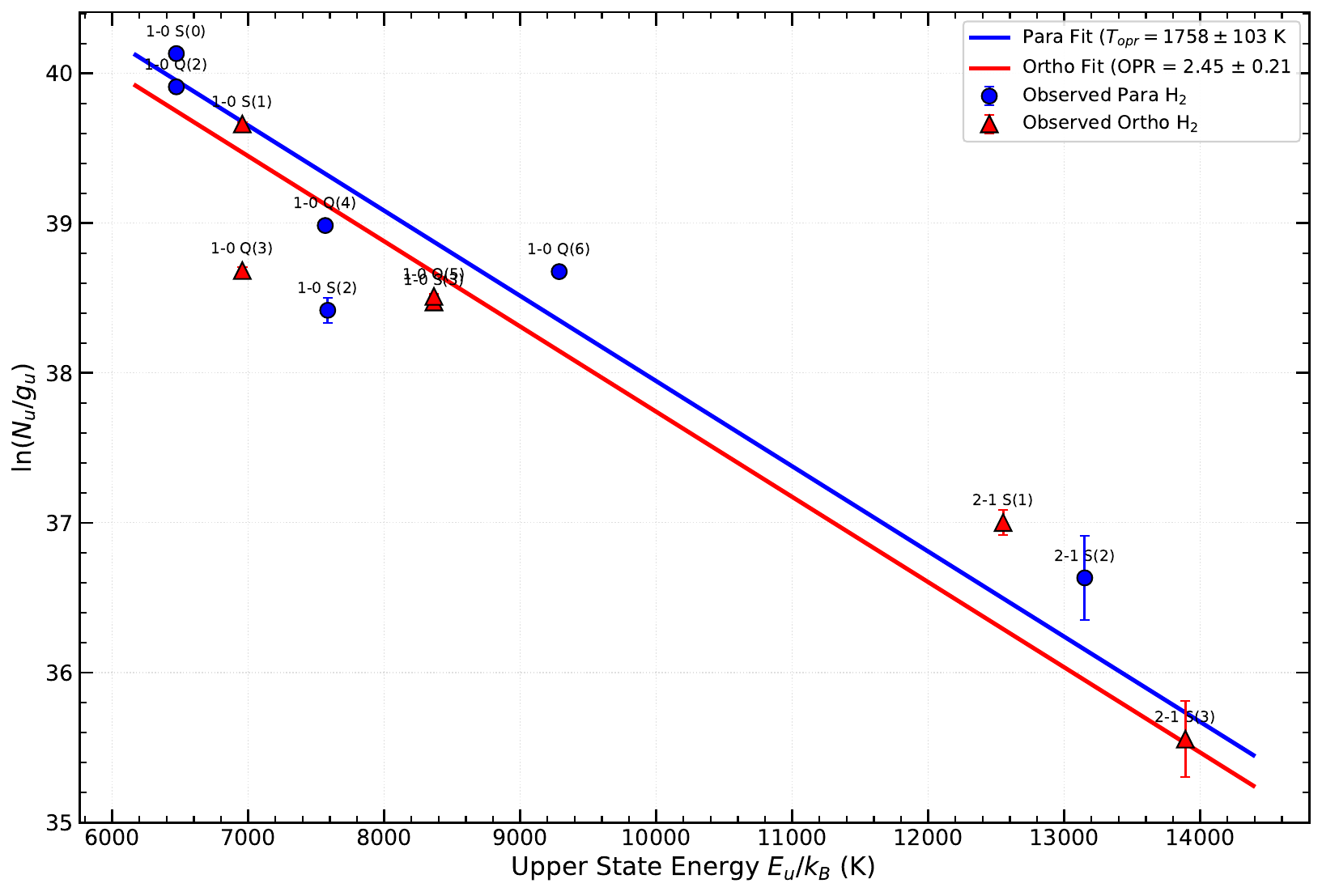}
\caption{Rovibrational diagram and single-temperature LTE model fit for $\mathrm{H_2}$ emission lines toward HOPS 370. The column densities normalized by statistical weights, $\ln(N_u/g_u)$, are plotted against the upper-state energies ($E_u/k_B$). Observed para-$\mathrm{H_2}$ ($g_s = 1$) and ortho-$\mathrm{H_2}$ ($g_s = 3$) lines are represented by blue circles and red triangles, respectively. Solid lines illustrate the simultaneous best-fit linear models for the para (blue) and ortho (red) populations. The fit yields a rovibrational temperature of $T_{\mathrm{opr}} = 1758 \pm 103$\,K and an ortho-to-para ratio of $\mathrm{OPR} = 2.45 \pm 0.21$.   The fit exhibits an RMS residual scatter of $\sigma_{\text{scatter}} \approx 0.39$ and a reduced chi-squared of $\chi^2_{\nu} \approx 13.64$. Individual $\mathrm{H_2}$ transitions are labeled textually above their respective data points.}
    \label{fig:h2_rotational_opr_diagram}
\end{figure*}

To determine the OPR toward HOPS~370, we constructed a rovibrational $\rm H_2$ excitation diagram, shown in Figure~\ref{fig:h2_rotational_opr_diagram}. The observed line fluxes were first corrected for interstellar extinction ($A_V = 23.24 \pm 1.25$\,mag). The dereddened fluxes were then converted into upper-level column densities ($N_u$) and normalized by their statistical weights, $g_u = g_s(2J_u+1)$, where $g_s=1$ and 3 correspond to para- and ortho-H$_2$, respectively.

Rather than assuming the high-temperature equilibrium value of $\rm OPR = 3$, we treated the OPR as a free parameter during the fitting procedure. In the excitation diagram, a non-equilibrium OPR produces a systematic vertical offset of $\ln({\rm OPR}/3)$ between the ortho- and para-H$_2$ transitions in $\ln(N_u/g_u)$ space. We therefore fitted the ortho and para populations simultaneously using a single-temperature Boltzmann model. In this approach, both spin isomers share a common slope, corresponding to a single rovibrational temperature, while the relative vertical displacement between the two sequences is determined by the OPR. The model parameters ($T_{\rm opr}$ and OPR) were obtained through a non-linear least-squares fit to all detected transitions, using the uncertainties in $\ln(N_u/g_u)$ as statistical weights.

The best-fitting model yields a rovibrational temperature of $T_{\rm opr}=1758 \pm 103$~K and an ortho-to-para ratio of ${\rm OPR}=2.45 \pm 0.21$. Although slightly below the LTE value of 3, the derived OPR indicates that the molecular gas is close to thermal equilibrium, consistent with excitation in warm, shock-heated gas associated with the protostellar outflow.

For comparison, the excitation temperature ($\rm T_{ex}$) derived in Section~\ref{sec:GlobalExcitationDiagram} was obtained from a global Boltzmann fit that assumes the equilibrium value $\rm OPR = 3$. That temperature is adopted for determining the total H$_2$ column density and partition function, whereas rovibrational temperature ($\rm T_{opr}$) from the present analysis  allows the OPR to vary freely in order to independently constrain the spin-state population distribution. The difference between the two fitted temperatures therefore reflects the distinct assumptions regarding the ortho and para level populations rather than evidence for multiple thermal components.

%%%%%%%%%%%%%%%%

\section{Discussion}\label{sec:discussions}
\subsection{Nature of the H$_2$ Emission and Relation to Protostellar Outflows}

The near-infrared K-band spectrum of HOPS\,370 is dominated by molecular hydrogen emission, including numerous S-branch and Q-branch transitions, indicating the presence of warm ($\sim$1500--3000 K) molecular gas. The strongest detected feature is the H$_2$ 1--0 S(1) line at 2.1218~$\mu$m, while the detection of multiple Q-branch lines implies a substantial column density of excited H$_2$ within the circumstellar environment. Such spectra are commonly observed toward embedded Class~0/I protostars associated with molecular outflows and shock-excited gas \citep{2001MNRAS.326..524D,2006A&A...449.1077C,2009A&A...496..153D}.

Molecular hydrogen has emerged as one of the most powerful tracers of protostellar feedback. Recent JWST observations of embedded protostars have demonstrated that H$_2$ emission can trace both highly collimated jets and wider-angle molecular winds, carrying a significant fraction of the mass and momentum injected into the surrounding cloud \citep{2023Natur.622...48R,2024A&A...691A.134C,2025A&A...695A.145V,2025ApJ...991...45D}. The observed properties of HOPS\,370 are broadly consistent with this interpretation. In contrast to many actively accreting Class~I protostars, the source exhibits weak Br$\gamma$ absorption and likely weak $\Hei$ P-cygni profiles. Since these atomic lines are generally associated with magnetospheric accretion \citep{1998AJ....116.2965M}, its weakness suggests that the K-band spectrum is not dominated by accretion-related processes. Instead, the observed emission appears to arise primarily from shock-excited molecular gas associated with ongoing outflow activity and the emitting gas is produced mainly in non-dissociative molecular shocks, likely magnetically mediated C-type shocks in which H$_2$ survives the shock passage and remains the principal coolant \citep{1980ApJ...241.1021D,1983ApJ...264..485D,1996ApJ...456..611K,2010MNRAS.406.1745F}.

This interpretation is further supported by the location of HOPS\,370 in the OMC-2 region of the Orion Molecular Cloud, a well-known site of active star formation and molecular outflows. HOPS\,370 is known to drive a powerful bipolar jet and molecular outflow. Using JWST/MIRI spectroscopy, \citet{2024ApJ...966L..22N} detected highly excited OH emission from shocked knots along the outflow and attributed it to ultraviolet radiation generated in fast shocks ($\sim$80 km\,s$^{-1}$) within the jet. More recently, \citet{2025ApJ...983..110T} showed that the H$_2$- and OH-emitting knots coincide with cavities and regions of reduced ice-column density, providing direct evidence that the outflow is influencing the surrounding envelope. Together, these observations support the conclusion that the rich H$_2$ spectrum of HOPS\,370 originates most likely in shock-heated molecular gas associated with its protostellar jet and outflow, rather than in the immediate accretion environment of the central source.

\subsection{Quantitative Excitation Diagnostics}

The H$_2$ line ratios provide a direct probe of the excitation mechanism operating in HOPS\,370. The measured 2--1 S(1)/1--0 S(1) ratio of $R=0.12$ is characteristic of thermal excitation and substantially lower than the values expected for pure ultraviolet fluorescence. Such low ratios are commonly observed in shock-excited molecular gas, where collisions efficiently thermalize the H$_2$ level populations.

The excitation temperature of $2078 \pm 215$ K and H$_2$ column density of $(1.93 \pm 0.93)\times10^{20}$ cm$^{-2}$, derived from the rovibrational transitions further supports this interpretation. Temperatures of a few thousand Kelvin are consistent with theoretical models of molecular shocks in dense protostellar environments, where H$_2$ acts as the dominant coolant of the post-shock gas \citep{1989ApJ...342..306H,1983ApJ...264..485D,1996ApJ...456..611K}. 

Comparing our derived line ratio ($R = 0.12 \pm 0.01$) and kinetic excitation temperature ($T_{\mathrm{ex}} = 2078 \pm 215$\,K) with theoretical shock models \citep[e.g.,][]{1996ApJ...456..611K, 2010MNRAS.406.1745F} provides a clear physical picture of the outflow dynamics. Non-dissociative $C$-type (continuous) shocks propagating at velocities $v_s \sim 15\text{--}30\,\mathrm{km\,s}^{-1}$ into dense envelope gas ($n_{\mathrm{H}_2} \sim 10^4\text{--}10^6\,\mathrm{cm}^{-3}$) maintain an extended ambipolar diffusion zone that produces warm, thermally excited molecular hydrogen without causing complete dissociation. In contrast, planar $J$-type shocks capable of reaching $T_{\mathrm{ex}} \approx 2100$\,K operate near the $\mathrm{H}_2$ dissociation limit ($v_s \gtrsim 20\,\mathrm{km\,s}^{-1}$), typically driving higher values of $R$ or causing complete molecular breakup. The combination of a low vibrational ratio ($R = 0.12$) and $T_{\mathrm{ex}} \approx 2080$\,K therefore points to $C$-type shocks as the dominant excitation mechanism along the outflow cavity walls of HOPS\,370.

The high rovibrational temperature captures the hot, localized shock fronts where the protostellar molecular jets interacts with  the surrounding ambient medium. The derived $\rm OPR$ of $2.45 \pm 0.21$, is below the high-temperature statistical equilibrium value of 3, indicating that the molecular hydrogen has not yet reached complete ortho--para equilibrium. Such non-equilibrium OPR values are commonly observed in shocked molecular gas and suggest that the spin-state populations retain a partial memory of an earlier, colder phase of the cloud. A progressive increase in the OPR toward the high-temperature equilibrium limit of 3 reflects the evolution of molecular gas from cold, quiescent reservoirs, where ortho-to-para conversion is suppressed by freeze-out at low temperatures, to shock-processed environments in which high-velocity shocks and reactive collisions efficiently re-equilibrate the molecular spin populations \citep[e.g.,][]{1998ApJ...498..246T,2006ApJ...649..816N,2025ApJ...995..199N}.
The measured value therefore implies that the molecular gas in HOPS\,370 has undergone substantial thermal processing by shocks but may not yet have fully erased the imprint of its colder pre-shock history.

Taken together, the measured $2\mbox{--}1\,\mathrm{S}(1)/1\mbox{--}0\,\mathrm{S}(1)$ ratio, excitation temperature, and OPR demonstrate that the molecular hydrogen emission in HOPS\,370 is thermally excited and is most likely associated with shocked gas in the protostellar outflow. The elevated, but sub-equilibrium, OPR indicates that the molecular gas has experienced substantial thermal processing and substantial ortho–para conversion, while still retaining a partial imprint of its colder pre-shock history. This is consistent with ongoing shock activity in which the molecular spin populations are evolving toward, but have not yet fully reached, the high-temperature equilibrium value.

\vspace{-2em}
\section{Summary and Conclusions}\label{sec:conclusions}

%\section{Conclusions}\label{sec:conclusions}

Using TANSPEC on the 3.6-m Devasthal Optical Telescope, we performed $K$-band ($1.94\text{--}2.52\,\mu\text{m}$) spectroscopy of the Orion Class\,I protostar HOPS\,370. Our key findings are:

\begin{enumerate}
    \item The spectrum is heavily dominated by $\rm H_2$ ro-vibrational transitions ($1\mbox{--}0$ S, $2\mbox{--}1$ S, and $1\mbox{--}0$ Q branches) with negligible atomic features, suggesting that near-infrared emission is most likely driven by protostellar outflow shocks.
    \item From the intrinsically linked $1\mbox{--}0$ Q(3) and $1\mbox{--}0$ S(1) lines, we derive a visual extinction of $A_V = 23.24 \pm 1.25$\,mag ($A_K = 2.50 \pm 0.13$\,mag), quantifying substantial circumstellar obscuration.
    \item The flux ratio $F(2\mbox{--}1\,{\rm S}(1))/F(1\mbox{--}0\,{\rm S}(1)) = 0.12 \pm 0.01$ and vibrational temperature $T_{\rm vib} = 2305 \pm 79$\,K rule out UV fluorescence, suggesting most likely  thermal shock excitation.
    \item A unified Boltzmann diagram across all dereddened lines yields a kinetic temperature of $T_{\rm ex} = 2078 \pm 215$\,K and an $\rm H_2$ column density of $(1.93 \pm 0.93) \times 10^{20}\text{ cm}^{-2}$.
    \item We derive an ortho-to-para ratio of $2.45 \pm 0.21$, indicating active thermal processing in warm shock fronts that has not yet reached full high-temperature equilibrium.
\end{enumerate}

\section*{Acknowledgements}
We thank the staff at the 1.3m DFOT and 3.6m DOT, Devasthal
(ARIES), for their cooperation during observations. It is a
pleasure to thank the members of 3.6m DOT team and IR
astronomy group at TIFR for their support during TANSPEC
observations. TIFR–ARIES Near Infrared Spectrometer
(TANSPEC) was built in collaboration with TIFR, ARIES,
and MKIR, Hawaii, for the DOT. K.S., D.K.O., J.P.N. and J.D. acknowledge the support of the Department of Atomic Energy, Government of India, under Project Identification No. RTI 4012. We are also thankful to Mr. Varghese Reji for his helpful guidance on the usage of the \texttt{pyTANSPEC} pipeline. 

\vspace{1em}
\noindent
\texttt{Facilities:} ARIES:DOT (TANSPEC), UKIDSS

\vspace{1em}
\noindent
\texttt{Software:} \texttt{pyTANSPEC} (v1.0) \citep{2026JApA...47...40R}, Astropy \citep{2013A&A...558A..33A}, Matplotlib \citep{2007CSE.....9...90H}

\vspace{-1em}

%%use \balance somewhere in the left column of the last page to balance the two columns in the end page

\bibliography{references}

%%References section
% \begin{theunbibliography}{}
% \vspace{-1.5em}
% \bibitem{latexcompanion}
% Clark D. H., Caswell J. L. 1976, MNRAS, 174, 267
% \bibitem{latexcompanion}
% Dickey, J. M., Salpeter, E. E., Terzian, Y. 1978, Astrophys. J. Suppl. Ser., 36, 77
% \bibitem{latexcompanion}
% Radhakrishnan, G. C. {\em et al.} 1980, in Evans A., Bode M. F., eds, Non-Solar Gamma Rays (COSPAR), Pergamon Press, Oxford, p. 163
% \bibitem{latexcompanion}
% Starrfield S., Iliadis C., Hix W. R. 2008, in Bode M. F., Evans A., eds, Classical Novae, 2nd edition, Cambridge University Press, Cambridge, p. 77
% \bibitem{latexcompanion}
% Van Loon J. Th. 2008, in Evans A. et al., eds, R S Ophiuchi (2006) and the Recurrent Nova Phenomenon, ASP Conference Series, Volume 401, p. 90
% \bibitem{latexcompanion}
% Zwicky, F. 1957, Morphological Astronomy, Springer-Verlag, Berlin, p. 258
% \end{theunbibliography}

\end{document}